\documentstyle[twocolumn,pra,aps,grafik]{revtex}

\begin{document}
\draft
\twocolumn[\hsize\textwidth\columnwidth\hsize\csname@twocolumnfalse\endcsname
\vspace{-0.5cm}
\begin{flushright}
Accepted to {\bf PHYSICAL REVIEW A} for publication
\end{flushright}
\title
{\bf Bose-Einstein condensation in a one-dimensional interacting 
system due to power-law trapping potentials}
\author{M. Bayindir, B. Tanatar, and Z. Gedik}
\address{
Department of Physics,
Bilkent University,
Bilkent, 06533 Ankara, Turkey}
\maketitle

\begin{abstract}
We examine the possibility of Bose-Einstein condensation in 
one-dimensional interacting Bose gas subjected to confining 
potentials of the form $V_{\rm ext}(x)=V_0(|x|/a)^\gamma$, 
in which $\gamma < 2$, 
by solving the Gross-Pitaevskii equation within the semi-classical 
two-fluid model. The condensate fraction, chemical potential, 
ground state energy, and specific heat of the system are 
calculated for various values of interaction strengths. Our 
results show that a significant fraction of the particles is
in the lowest energy state for finite number of particles at low 
temperature indicating a phase transition for weakly
interacting systems.
\end{abstract}

\vskip1pc]
\narrowtext
\newpage

\section{Introduction}

The recent observations of Bose-Einstein condensation 
(BEC) in trapped atomic gases\cite{1,2,3,4,5} have renewed 
interest in bosonic systems\cite{6,7}. BEC is characterized 
by a macroscopic occupation of the ground state for $T<T_0$,
where $T_0$ depends on the system parameters. The success of
experimental manipulation of externally applied trap potentials
bring about the possibility of examining two or even
one-dimensional Bose-Einstein condensates.
Since the transition temperature $T_0$ increases with decreasing 
system dimension, it was suggested that BEC may be achieved more 
favorably in low-dimensional systems\cite{8}.
The possibility of BEC in one - (1D) and two-dimensional (2D) 
homogeneous Bose gases is ruled out by the Hohenberg 
theorem\cite{9}. However, 
due to spatially varying potentials which break the translational 
invariance, BEC can occur in low-dimensional inhomogeneous systems. 
The existence of BEC is shown in a 1D noninteracting Bose gas in the 
presence of a gravitational field\cite{10}, an 
attractive-$\delta$ impurity\cite{11}, and power-law 
trapping potentials\cite{12}. Recently, many authors have discussed the 
possibility of BEC in 1D trapped Bose 
gases relevant to the magnetically trapped ultracold
alkali-metal atoms\cite{13,14,15,16,17,18}.
Pearson and his co-workers\cite{19} 
studied the interacting Bose gas in 1D power-law potentials employing
the path-integral Monte Carlo (PIMC) method. They have found
that a macroscopically large number of atoms occupy the lowest
single-particle state in a finite system of hard-core bosons at
some critical temperature.
It is important to note that the recent BEC experiments are carried 
out with finite number of atoms (ranging from several thousands to 
several millions), therefore the thermodynamic limit argument in 
some theoretical studies\cite{15} does not apply here\cite{8}.  

The aim of this paper is to study the two-body interaction effects 
on the BEC in 1D systems under power-law trap potentials. For ideal
bosons in harmonic oscillator traps transition to a condensed
state is prohibited. It is anticipated that the external potentials
more confining than the harmonic oscillator type would be
possible experimentally.
It was also argued\cite{15} that in the thermodynamic limit there 
can be no BEC phase transition for nonideal bosons in 1D.
Since the realistic systems are weakly interacting and contain
finite number of particles, we employ the
mean-field theory\cite{20,21} as applied to a two-fluid model. Such
an approach has been shown to capture the essential physics in
3D systems\cite{21}. The 2D version\cite{22} is also in qualitative
agreement with the results of PIMC simulations on hard-core
bosons\cite{23}. In the remaining sections we outline the two-fluid 
model and present our results for an interacting 1D Bose gas in 
power-law potentials.

\section{Theory}

In this paper we shall investigate the Bose-Einstein condensation 
phenomenon for 1D interacting Bose gas confined in a 
power-law potential:
\begin{equation}
V_{\rm ext}(x)=V_0 \left(\frac{|x|}{a}\right)^\gamma \; ,
\end{equation}
where $V_0$ and $a$ are some suitable energy and length
parameters defining the external potential, and $\gamma$
controls the confinement strength. Presumably, they can be
experimentally adjusted.
Using the semi-classical density of states, the transition temperature 
$T_0$ and the fraction of condensed particles $N_0/N$ for 
the noninteracting system were calculated as\cite{12}
\begin{equation}
k_BT_0= \left[ \frac{N}{\kappa F(\gamma)G(\gamma)}\right
]^{2\gamma/(2+\gamma)} \; ,
\end{equation}
and
\begin{equation}
N_0/N=1-\left({T \over T_0 }\right)^{1/\gamma+1/2}\; ,
\end{equation}
where $\kappa=2(2m)^{1/2}a/\gamma hV_0^{1/\gamma}$ ($m$ is the
mass of bosons and $h$ is the Planck's constant), and
\begin{equation}
F(\gamma)=\int_0^1\frac{x^{1/\gamma-1}\,  dx}{\sqrt{1-x}}\; ,
\end{equation}
and
\begin{equation}
G(\gamma)=\int_0^\infty \frac{x^{1/\gamma-1/2} \; dx}
{e^x-1}=\Gamma(1/\gamma+1/2) \; \zeta(1/\gamma+1/2) \; , 
\end{equation}
in which $\Gamma(x)$ and $\zeta(x)$ are the gamma and the Riemann 
zeta-functions, respectively. The total energy of the noninteracting
system for 
$T<T_0$ ($\mu=0$) is given by
\begin{equation}
{{\langle E\rangle} \over N k_BT_0}={ \Gamma(1/\gamma+3/2) 
\;\zeta(1/\gamma+3/2)  \over \Gamma(1/\gamma+1/2) \;\zeta(1/\gamma+1/2)
}  \left( {T\over T_0}\right)^{1/\gamma +3/2}\; .
\end{equation}
Figure 1 shows the variation of the critical temperature $T_0$ as
a function of the exponent $\gamma$ in the trapping potential. It 
should be noted that $T_0$ vanishes for harmonic potential 
due to the divergence of the function $G(\gamma=2)$. It appears
that the maximum $T_0$ is attained for $\gamma\approx 0.5$, and
for a constant trap potential (i.e. $V_{\rm ext}(x)=V_0$) the BEC
disappears consistent with the Hohenberg theorem.

\bfig{h}\ff{0.45}{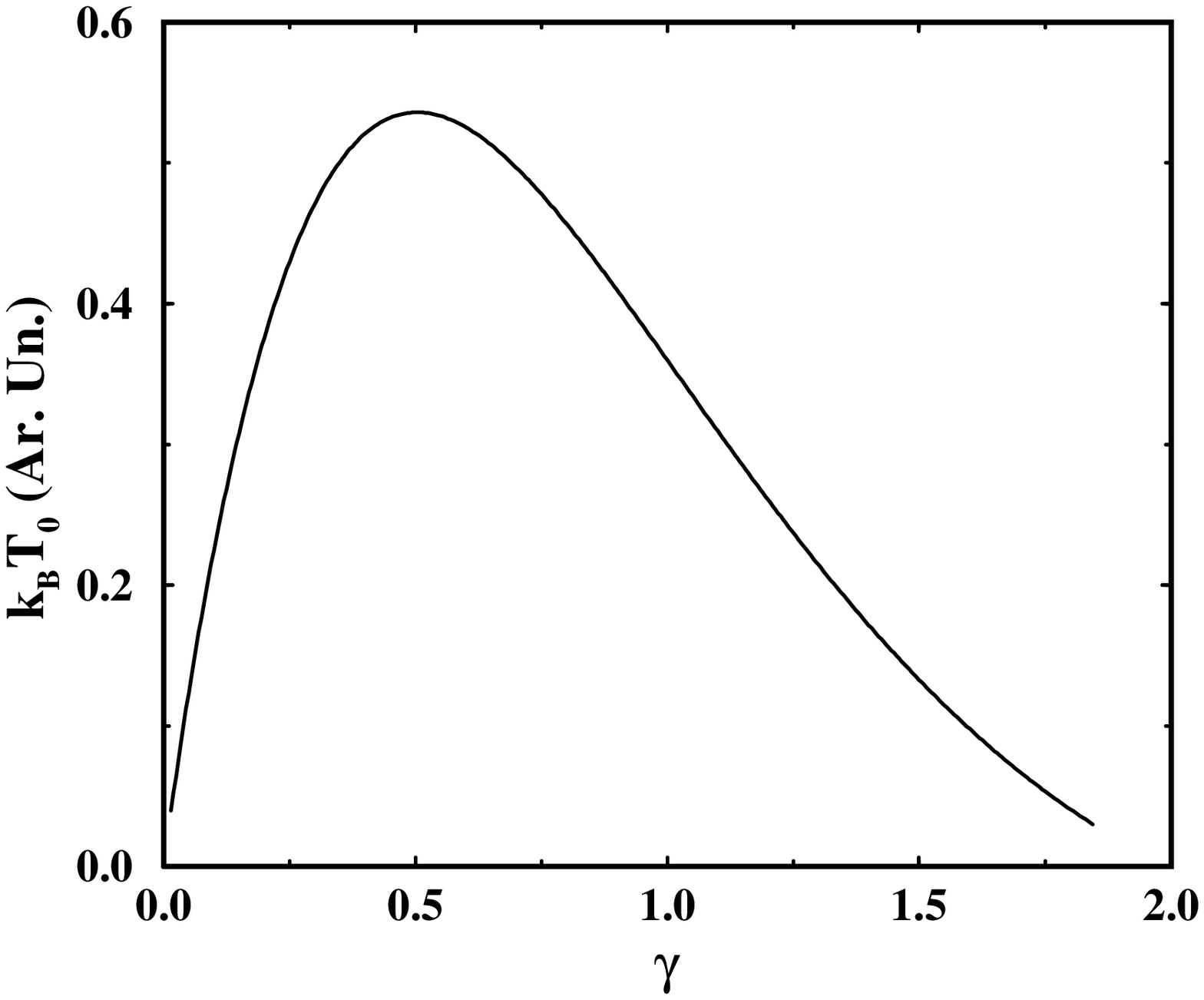}\efig{The variation of the critical 
temperature $T_0$ with the external potential exponent $\gamma$.}{f1}

We are interested in how the short-range interaction effects
modify the picture presented above. To this end, we employ the
mean-field formalism and describe the collective dynamics of a
Bose condensate by its macroscopic time-dependent wave function 
$\Upsilon(x,t)=\Psi(x)\exp{(-i\mu t)}$, where $\mu$ is the chemical 
potential.
The condensate wave function $\Psi(x)$ satisfies the 
Gross-Pitaevskii (GP) equation\cite{24,25}
\begin{equation}
\left[ -{\hbar^2\over 2m}{d^2\over dx^2}+V_{\rm ext}(x)+2gn_1(x) +g 
\Psi^2 (x)\right] \Psi(x)=\mu\Psi(x)\; ,
\end{equation}
where $g$ is the repulsive, short-range interaction strength, 
and $n_1(x)$ is the average noncondensed particle distribution 
function. We treat the interaction strength $g$ as a
phenomenological parameter without going into the details of
actually relating it to any microscopic description\cite{26}. 
In the semi-classical two-fluid model\cite{27,28} 
the noncondensed particles can be treated as bosons in an 
effective potential\cite{21,29}
\begin{equation}
V_{\rm eff}(x)=V_{\rm ext}(x)+2gn_1(x)+2g\Psi^2(x)\; .
\end{equation}
The density distribution function is given by
\begin{equation}
n_1(x)=\int {dp\over 2\pi\hbar}\; {1\over\exp{\{[p^2/2m+
V_{\rm eff}(x)-\mu]/k_BT\}}-1}\; ,
\end{equation}
and the total number of particles $N$ fixes the chemical potential
through the relation
\begin{equation}
N=N_0+\int {\rho(E)\; dE\over \exp{[(E-\mu)/k_BT]}-1}\; ,
\end{equation}
where $N_0=\int \Psi^2(x)\,dx$ is the number of condensed particles, 
and the semi-classical density of states is determined by
\begin{equation}
\rho(E)={\sqrt{2m}\over h}\,\int_{V_{\rm eff}(x)<E}{ dx \over 
\sqrt{E-V_{\rm eff}(x)}}\; .
\end{equation}
The GP equation yields a simple solution when the kinetic 
energy term is neglected (the Thomas-Fermi approximation)
\begin{equation}
\Psi^2(x)={\mu-V_{\rm ext}(x)-2gn_1(x) \over g} \theta 
[ \mu-V_{\rm ext}(x)-2gn_1(x) ]\; ,
\end{equation}
where $\theta[x]$ is the unit step function. More precisely,
the Thomas-Fermi approximation\cite{7,20,30} would be valid when the
interaction energy $\sim gN_0/\Lambda$, far exceeds the kinetic
energy $\hbar^2/2m\Lambda^2$, where $\Lambda$ is the spatial
extent of the condensate cloud. For a linear trap potential
(i.e. $\gamma=1$), a variational estimate for $\Lambda$ is given
by
$\Lambda=\left[\hbar^2/2m\,(\pi/2)^{1/2}\,2a/V_0\right]^{1/3}$.
We note that the Thomas-Fermi approximation would breakdown for
temperatures close to $T_0$ where $N_0$ is expected to become
very small.

The above set of equations [Eqs.\,(9)-(12)] need to be solved
self-consistently to obtain the various physical quantities such
as the chemical potential $\mu(N,T)$, the condensate fraction
$N_0/N$, and the effective potential $V_{\rm eff}$. In a 3D
system, Minguzzi {\it et al}.\cite{21} solved a similar
system of equations numerically and also introduced an
approximate semi-analytical solution by treating the
interaction effects perturbatively. Motivated by the 
success\cite{21,22} of the perturbative approach we consider 
a weakly interacting system in 1D.
To zero-order in $gn_1(r)$, the effective potential becomes
\begin{equation}
V_{\rm eff}(x)=\left\{  \begin{array}{ll}
V_{\rm ext}(x) & \quad\hbox{if}\quad \mu< V_{\rm ext}(x) \, \\
2\mu-V_{\rm ext}(x)& \quad\hbox{if}\quad \mu>  V_{\rm ext}(x)\; .
\end{array}
\right.
\end{equation}
Figure 2 displays the typical form of the effective potential 
within our semi-analytic approximation scheme. The most
noteworthy aspect is that the effective potential as seen by the
bosons acquire a double-well shape because of the interactions.
We can explain this result by a simple argument. Let the number
of particles in the left and right wells be $N_L$ and $N_R$,
respectively, so that $N=N_L+N_R$. The nonlinear or interaction
term in the GP equation may be approximately regarded as
$V=N_L^2+N_R^2$. Therefore, the problem reduces to the
minimization of the interaction potential $V$, which is achieved
for $N_L=N_R$.

\bfig{h}\ff{0.45}{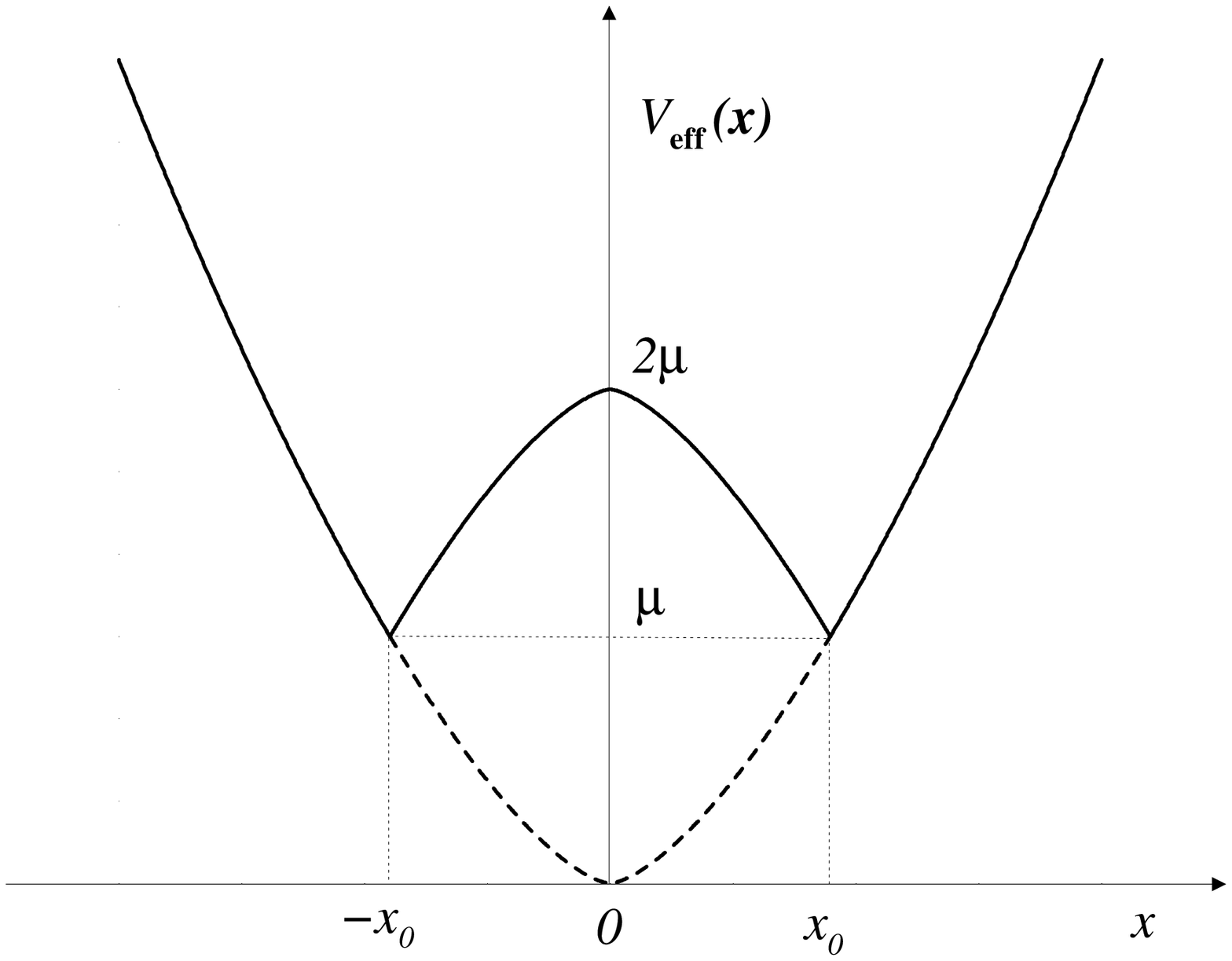}\efig{Effective potential $V_{\rm eff}(x)$ 
in the presence of interaction ($x_0=(\mu/V_0)^{1/\gamma}a$). Thick 
dotted line  represents external potential $V_{\rm ext}(x)$.}{f2}

The number of condensed atoms is calculated to be
\begin{equation}
N_0={2\gamma a  \over (1+\gamma) g V_0^{1/\gamma}}\; 
\mu^{1/\gamma+1}\; .
\end{equation}
The density of states is given by
\begin{equation}
\rho(E)= \kappa \left\{ 
\begin{array}{ll}
H(\gamma, \mu, E)\; (2\mu-E)^{1/\gamma-1/2}  & 
\quad\hbox{if}\quad \mu<E<2\mu\, \\
F(\gamma)\; E^{1/\gamma-1/2}   & \quad\hbox{if}\quad E>2\mu\, , 
\end{array}
\right.
\end{equation}
where
\[ H(\gamma, \mu, E)=\int_1^{E/(2\mu-E)} \frac{x^{1/\gamma-1} \, dx}
{\sqrt{x-1}}\; .\] 
Using the above density of states, conservation of total number of 
particles gives us a transcendental equation for the chemical potential
\begin{equation}
N=N_0+ \kappa \; 
(k_BT)^{1/\gamma+1/2}\; I(\gamma, \mu, T)\; ,
\label{eq:fraction}
\end{equation}
where 
\begin{eqnarray*}
I(\gamma, &&\mu, T)=F(\gamma) \; \int_{2\mu/k_BT}^\infty \; 
\frac{x^{1/\gamma-1/2} \,dx} {z e^x -1} \\
&& +\int_{\mu/k_BT}^{2\mu/k_BT} H(\gamma, \mu, xk_BT) \; 
\frac{(2\mu/k_BT-x)^{1/\gamma-1/2} \, dx}
{z e^x -1}\; . 
\end{eqnarray*}
in which $z=e^{-\mu/k_BT}$. The chemical potential $\mu(N,T)$ is 
determined from the solution of 
Eq.\,(16). Finally, the total energy of the interacting
system can be written as
\begin{equation}
\langle E\rangle=[{\langle E\rangle}_{\rm nc}(N-N_0)/2+{\langle 
E\rangle}_{\rm c}]/N\; ,
\end{equation}
where ${\langle E\rangle}_{\rm nc}$ is the energy of the 
noncondensed particles
\begin{eqnarray}{\langle E\rangle}_{\rm nc}&=&\int {E \rho(E)\; dE\over 
\exp{[(E-\mu)/k_BT]}-1}\nonumber \\
&=& \kappa \; (k_BT)^{1/ \gamma+1/2}\; J(\gamma, \mu, T) \;,
\end{eqnarray}
where
\begin{eqnarray*}
&&J(\gamma, \mu, T)= \int_{2\mu/k_BT}^\infty \; \frac{x^{1/
\gamma+1/2} \,dx} {z e^x-1} \\
&& +\int_{\mu/k_BT}^{2\mu/k_BT} H(\gamma, \mu, x) \; 
\frac{(2\mu/k_BT-x)^{1/\gamma+1/2} \, dx}{z e^x-1}\; . 
\end{eqnarray*} 
and  ${\langle E\rangle}_{\rm c}$ is the energy of the particles
in the condensate
\begin{equation}
{\langle E\rangle}_{\rm c}={g \over 2} \int \Psi^4(x)\;dx={2a
\gamma^2\mu^{2+1/\gamma} \over (1+\gamma)(2\gamma+1)gV_0^{1/\gamma}}\,
.
\end{equation}
The kinetic energy of the condensed particles
is neglected within our Thomas-Fermi approximation to the GP
equation. 

\section{Results and discussion}

Up to now we have based our formulation for
arbitrary $\gamma$, but in the rest 
of this work we shall present our results for $\gamma=1$. Our
calculations show that the results for other values of $\gamma$
are qualitatively similar.
In Figs. 3 and 4 we calculate the condensate 
fraction as a function of temperature for various values of the
interaction strength $\eta=g/V_0a$ (at constant $N=10^5$) and 
different number of particles 
(at constant $\eta=0.001$), respectively. We observe that as the
interaction strength $\eta$ is increased, the depletion of the
condensate becomes more appreciable (Fig.\,3). As shown in the 
corresponding figures, a significant fraction of the particles 
occupies the ground state of the system for $T < T_0$.
The temperature dependence of the chemical potential is plotted in 
Figs.\,5 and 6 for various interaction strengths (constant $N=10^5$) 
and different number of particles (constant $\eta=0.001$) respectively.

\bfig{h}\ff{0.45}{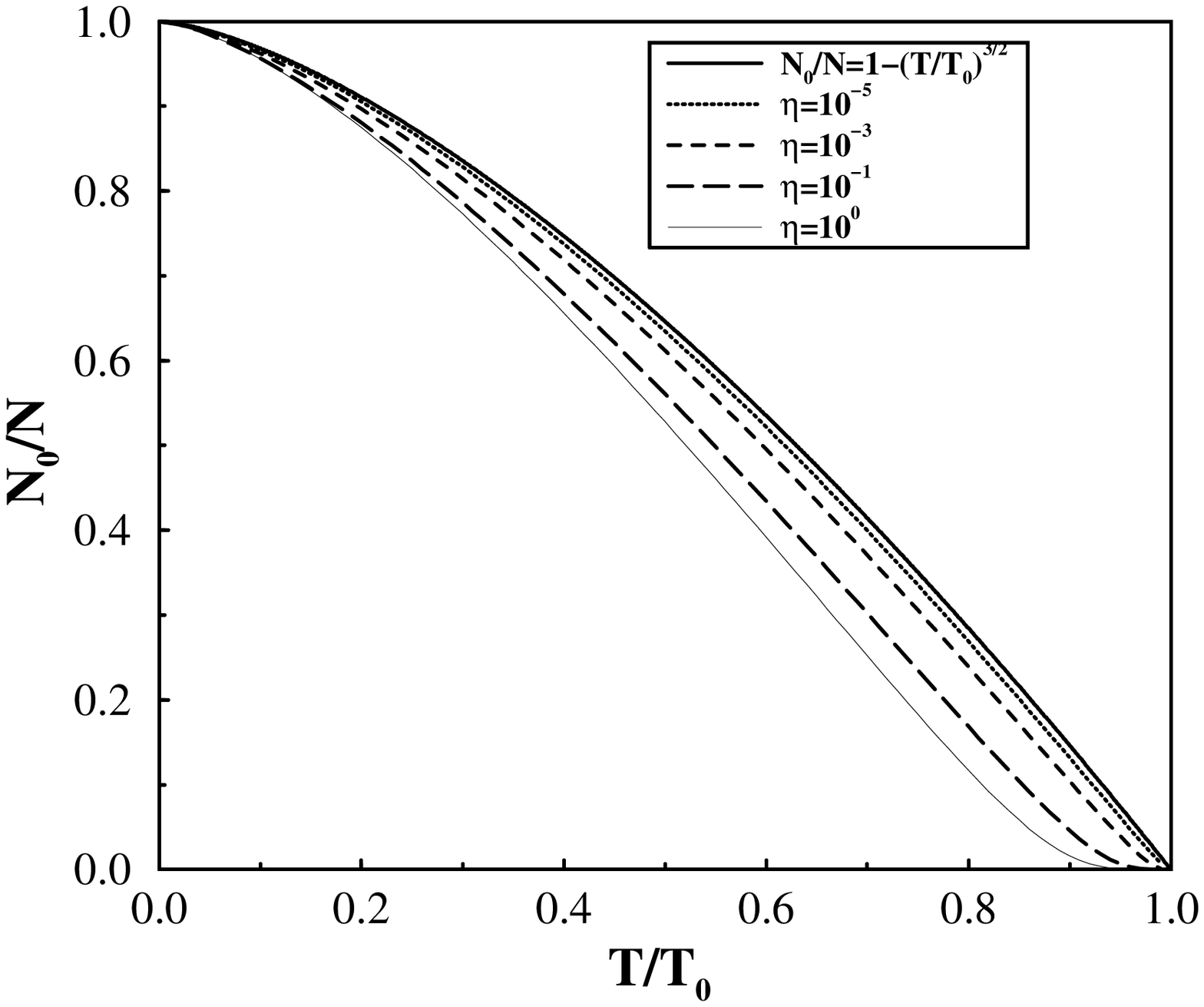}\efig{The condensate fraction $N_0/N$ 
versus temperature $T/T_0$ for $N=10^5$ and for various interaction 
strengths $\eta$.}{f3}

Effects of interactions on $\mu(N,T)$ are seen as large
deviations from the noninteracting behavior for $T<T_0$. 
In Fig.\,7 we show the ground state energy of an interacting 1D
system of bosons as a function of temperature for different
interaction strengths. For small $\eta$, and $T<T_0$, $\langle
E\rangle$ is similar to that in a noninteracting system. As
$\eta$ increases, some differences start to become noticeable,
and for $\eta\approx 1$ we observe a small bump developing in
$\langle E\rangle$. This may indicate the breakdown of our
approximate scheme for large enough interaction strengths, as we
can find no fundamental reason for such behavior. It is also
possible that the Thomas-Fermi approximation employed 
is violated as the transition to a condensed state is
approached.

\bfig{h}\ff{0.45}{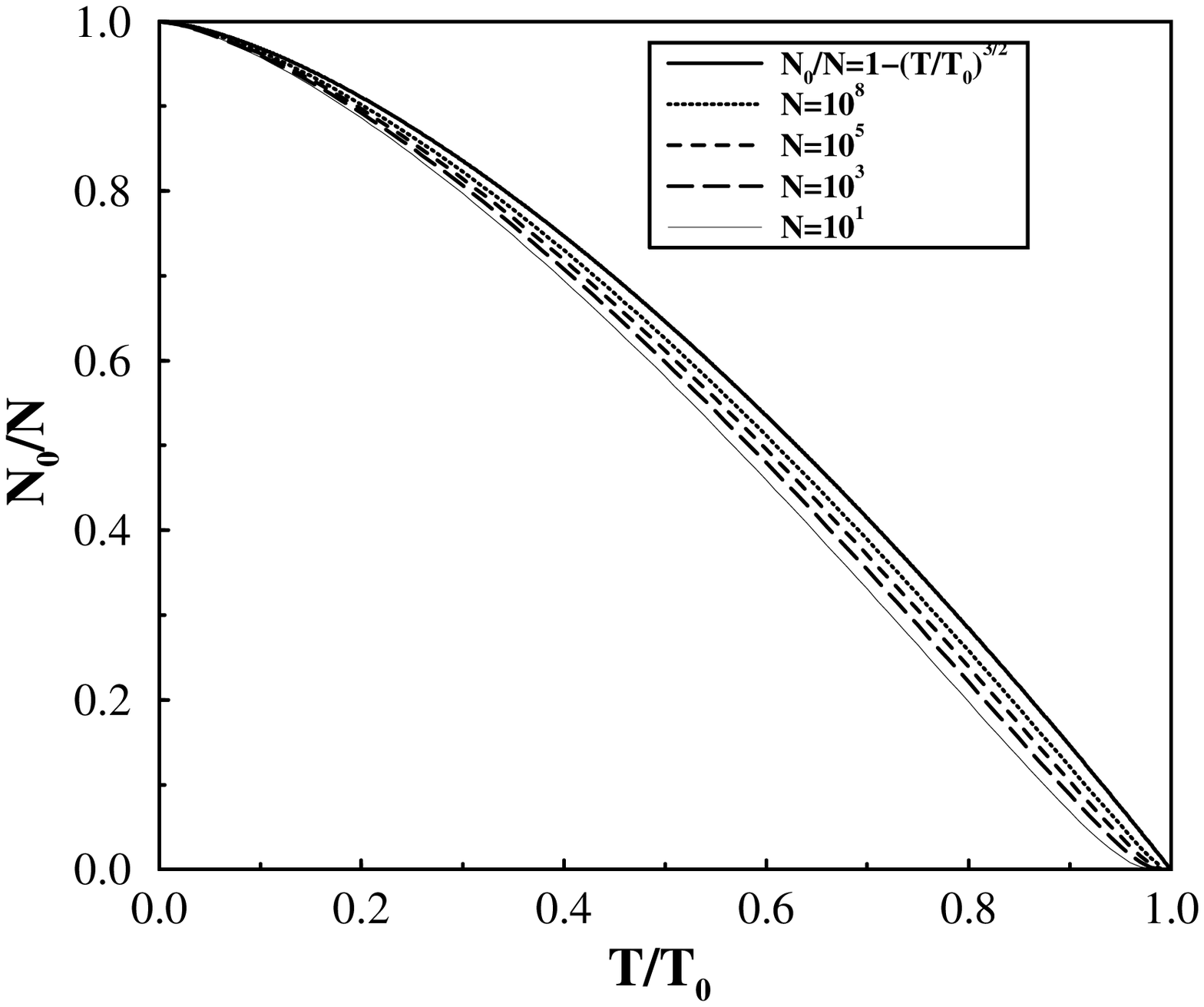}\efig{The condensed fraction $N_0/N$ 
versus temperature $T/T_0$ for $\eta=0.001$ and for different 
number of particles $N$.}{f4}

\bfig{h}\ff{0.45}{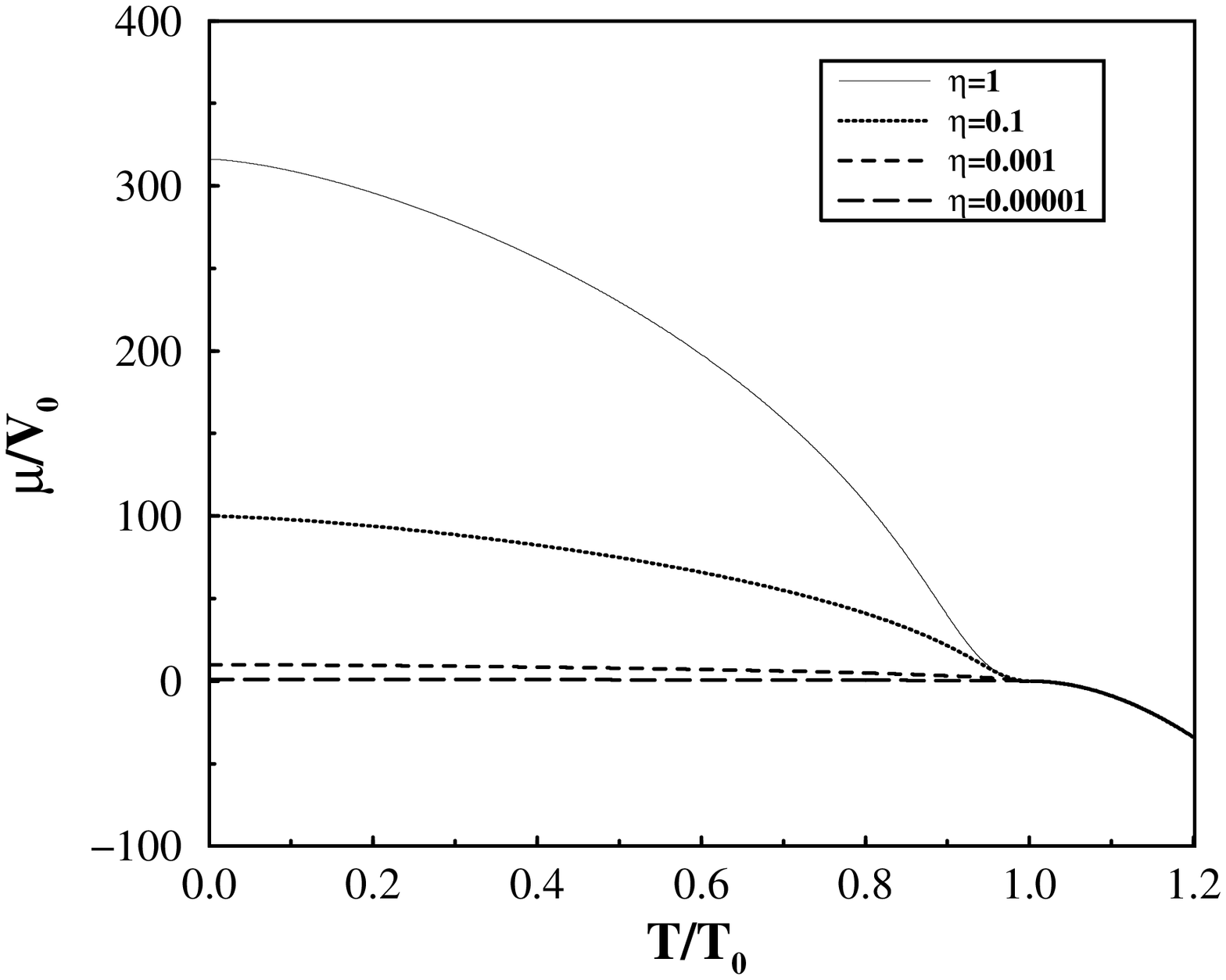}\efig{The temperature dependence of the 
chemical potential $\mu(N,T)$ for various interaction strength and 
for $N=10^5$ particles.}{f5}

Although it is
conceivable to imagine the full solution of the mean-field
equations [Eq.\,(9)-(12)] may remedy the situation for larger
values of $\eta$, the PIMC 
simulations\cite{19} also seem to indicate that the condensation
is inhibited for strongly interacting systems. 
The results
for the specific heat calculated from the total energy curves, 
i.e. $C_V=d\langle E\rangle/dT$, are depicted in Fig.\,8. The
sharp peak at $T=T_0$ tends to be smoothed out with increasing
interaction strength. It is known that the effects of finite number
of particles are also responsible for such a behavior\cite{20}. 
In our treatment these two effects are not disentangled. It was
pointed out by Ingold and Lambrecht\cite{14} that the identification
of the BEC should also be based on the behavior of $C_V$ around
$T\approx T_0$.
\bfig{h}\ff{0.45}{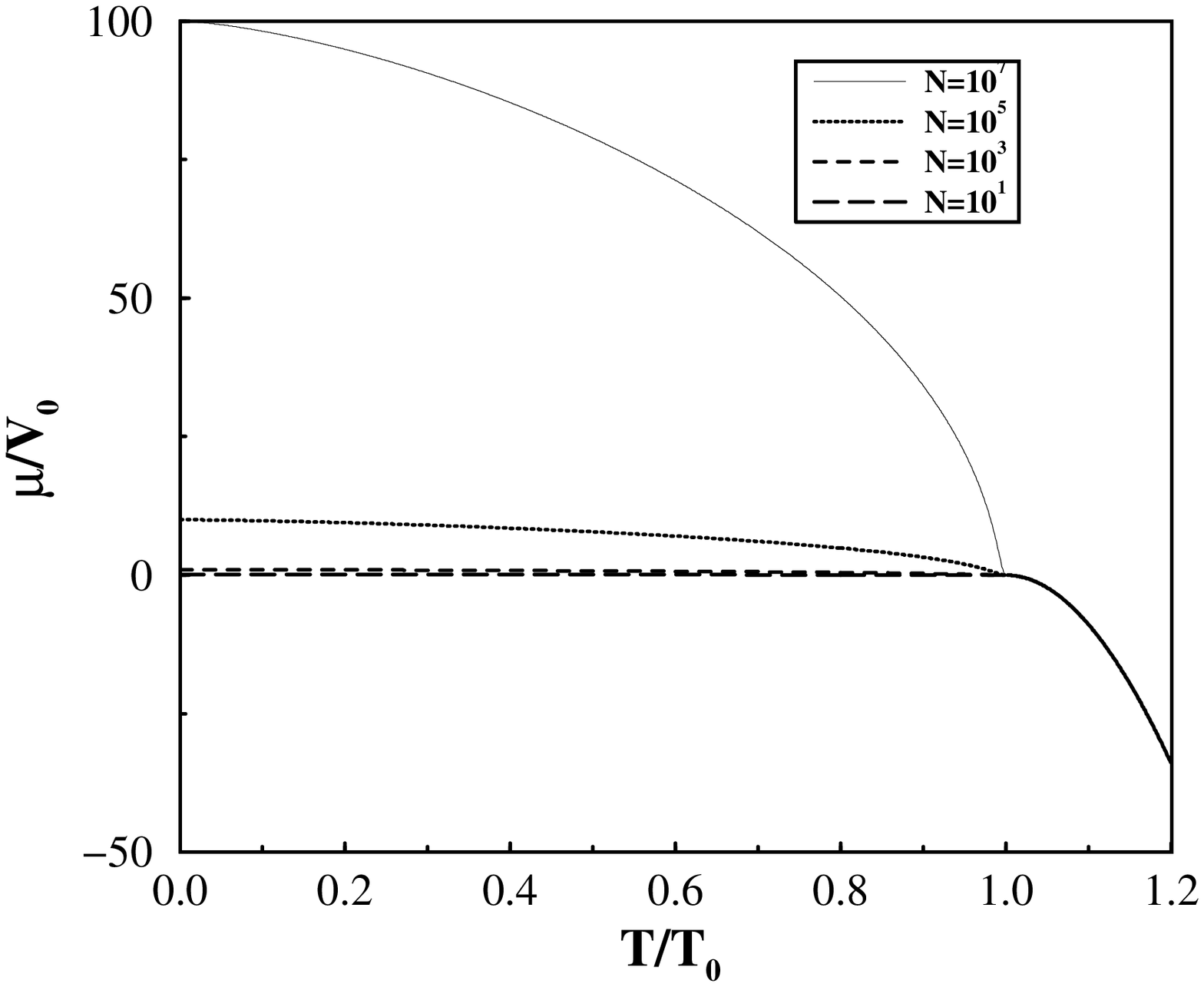}\efig{The temperature dependence of the 
chemical potential $\mu(N,T)$ for different number of particles $N$
 and for $\eta=0.001$.}{f6}
\bfig{h}\ff{0.45}{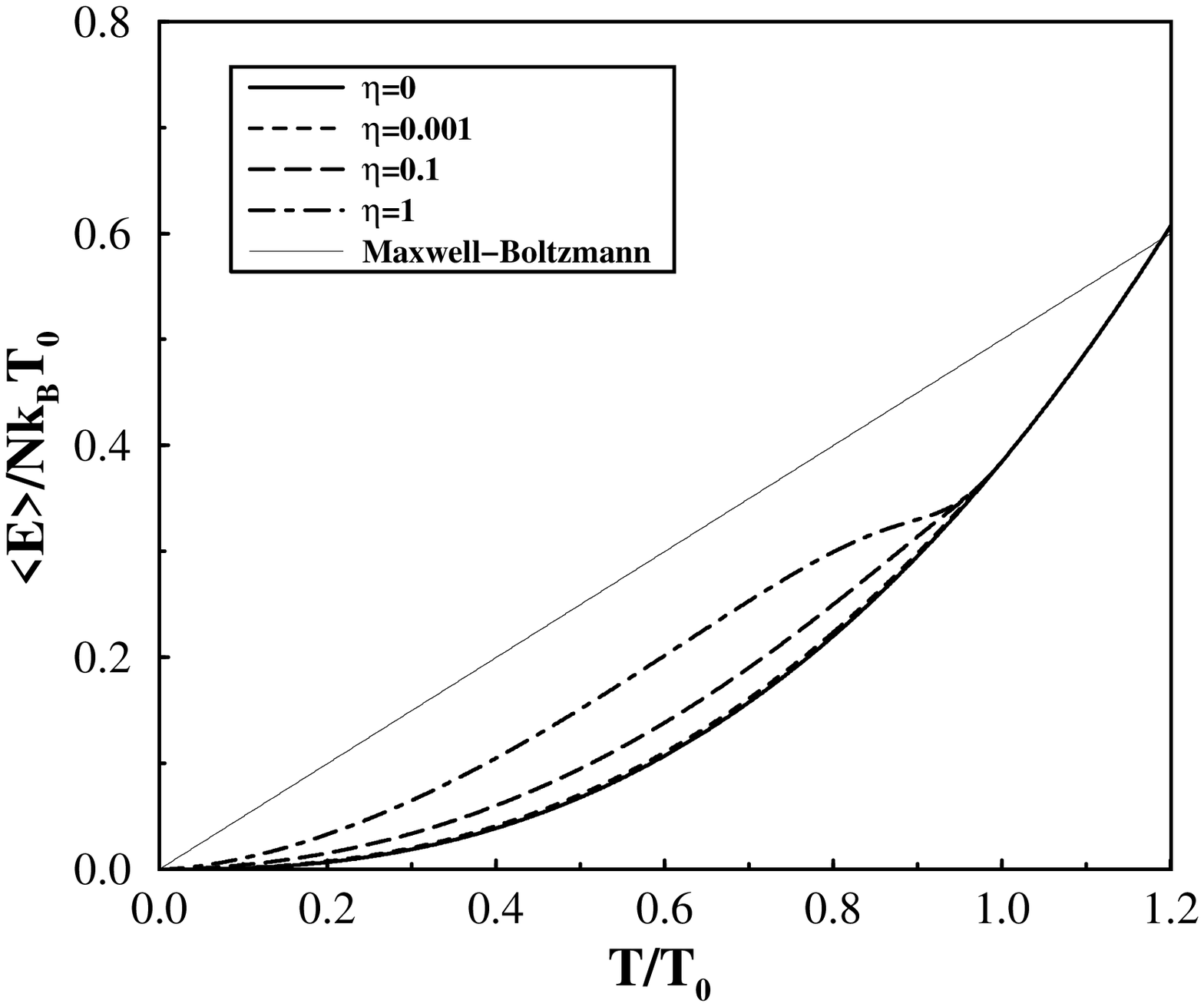}\efig{The temperature dependence of the total
 energy of 1D Bose gas for various interaction strengths $\eta$ and 
$N=10^5$ particles.}{f7}
Our calculations indicate that the peak
structure of $C_V$ remains even in the presence of weak
interactions, thus we are led to conclude that a true transition
to a Bose-Einstein condensed state is predicted within the
present approach.

\bfig{t}\ff{0.45}{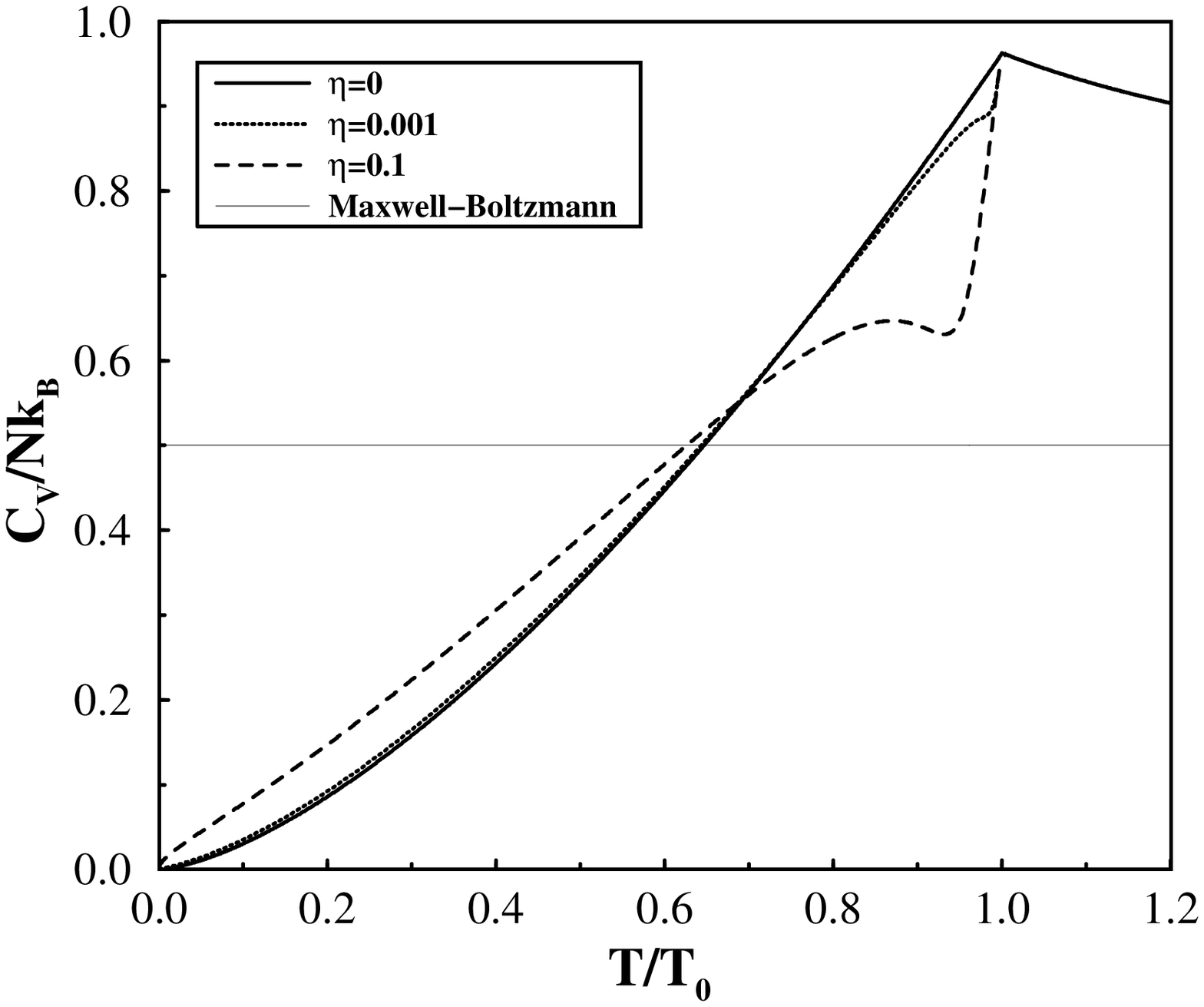}\efig{The temperature dependence of 
the specific heat $C_V$ for various interaction strengths $\eta$ 
and $N=10^5$ particles.}{f8}

\section{Concluding remarks}

In this work we have applied the mean-field, semi-classical
two-fluid model to interacting bosons in 1D power-law trap
potentials. We
have found that for a range of interaction strengths the
behavior of the thermodynamic quantities resembles to that of 
non-interacting bosons. Thus, BEC in the sense of macroscopic
occupation of the ground state, occurs when the short-range
interparticle interactions are not too strong. Our results
are in qualitative agreement with the recent PIMC 
simulations\cite{19} of similar systems. Both 2D and 1D
simulation results\cite{19,23} indicate a phase transition
for a finite number system, in contrast to the situation in the
thermodynamic limit. Since systems of much larger size can be
studied within the present approach, our work complements the PIMC
calculations. 

The possibility of studying the tunneling phenomenon of condensed 
bosons in spatially different regions separated by a barrier has 
recently attracted some attention\cite{31,32,33,34}. 
In particular, Dalfovo {\it et al}.\cite{32} have shown that a 
Josephson-type
tunneling current may exist for bosons under the influence of a
double-well trap potential. Zapata {\it et al}.\cite{34} have
estimated the Josephson coupling energy in terms of the
condensate density. It is interesting to speculate on
such a possibility in the present case, since the effective 
potential in our description is of
the form of a double-well potential (cf. Fig.\,2). In our
treatment, the interaction effects modify the single-well trap
potential into one which exhibits two minima. Thus if we think
of this effective potential as the one seen by the condensed
bosons and according to
the general arguments\cite{31,32,33,34} based on two weakly 
connected systems we should have an oscillating flux of
particles when the chemical potential in the two wells is
different. Any configuration with $N_L\ne N_R$ which is always
the case for odd number of bosons will result in an oscillatory
motion. It would be interesting to explore these ideas in future
work.

\acknowledgements{
This work was supported by the Scientific and Technical 
Research Council of Turkey (TUBITAK) under Grant No. TBAG-1736
and TBAG-1662. We gratefully acknowledge useful discussions with 
Prof. C. Yalab{\i}k and E. Demirel.}

\end{document}